\documentclass[12pt]{article}
\usepackage{epsfig}
\usepackage{amssymb,amsmath}
\usepackage{slashed}
\usepackage{multirow}

\newcommand{\bea}{\begin{eqnarray}}
\newcommand{\eea}{\end{eqnarray}}
\newcommand{\beq}{\begin{equation}}
\newcommand{\eeq}{\end{equation}}

\makeatletter
\def\alt{\mathrel{\mathpalette\gl@align<}}
\def\agt{\mathrel{\mathpalette\gl@align>}}
\def\gl@align#1#2{\lower.6ex\vbox{\baselineskip\z@skip\lineskip\z@
\ialign{$\m@th#1\hfil##\hfil$\crcr#2\crcr\sim\crcr}}} \makeatother

\begin{document}

\begin{center}
\baselineskip 20pt {\Large\bf  Simplified Smooth Inflation
with Observable Gravity Waves} \vspace{1cm}

Mansoor Ur Rehman$^{\star}$
and Qaisar Shafi$^{\dagger}$
\\[1mm]
\end{center}
\vspace*{0.50cm}
\centerline{$^{\star}$ \it
 Department of Physics, University of Basel,}
\centerline{\it
Klingelbergstr.~82, CH-4056 Basel, Switzerland,}
\vspace*{0.2cm}
\centerline{$^{\dagger}$ \it
Bartol Research Institute, Department of Physics and Astronomy,}
\centerline{\it
University of Delaware, Newark, Delaware 19716, USA}
\vspace*{0.50cm}

\begin{abstract}
We consider a simplified version of supersymmetric smooth hybrid inflation which 
contains a single ultraviolet cutoff $m_P = 2.4 \times 10^{18}$ GeV, instead of the two cutoffs
$m_P$ and $M_* \sim $ few $\times 10^{17}$ GeV that are normally employed.
With global supersymmetry the scalar spectral index $n_s \simeq 0.97$, which is in very good agreement with the WMAP observations. With a non-minimal K\"ahler potential, the supergravity version of the model is compatible with the current central values of $n_s$ and also yields potentially observable gravity waves (tensor to scalar ratio $r \lesssim 0.02$).
\end{abstract}

\date{}

  Supersymmetric (SUSY) hybrid inflation models \cite{Dvali:1994ms,Copeland:1994vg}, provide
an interesting possibility of realizing inflation in the grand unified theories (GUTs) of
particle physics 
\cite{Senoguz:2003zw}-\cite{Antusch:2010va}. 
Among its attractive feature are the solution of eta problem,
adequately suppressed supergravity (SUGRA) corrections and consistency with  the recent
WMAP7 data \cite{Komatsu:2010fb}. In the standard version of susy hybrid inflation, gauge symmetry is usually
broken at the end of inflation. This implies that the topological defects
(such as monopoles), if present, are produced after the inflation and
their presence is in contradiction with the experimental observations.
In order to solve this problem of topological defects, various extensions of
susy hybrid inflation have been proposed. Among all these variants,
shifted \cite{Jeannerot:2000sv} and smooth \cite{Lazarides:1995vr}
hybrid inflation models are the simplest ones.
In these models inflation occurs along the `shifted'
tracks where gauge symmetry is broken during inflation.
This then solves the problem of topological defects by
inflating them away and by reducing their density under
the observational limits. However, in contrast to shifted hybrid inflation,
in the smooth hybrid inflation scenario inflation ends smoothly 
without any water fall effect.

  In this brief report we will consider a simplified version of
smooth hybrid inflation model. By simplified we mean that the
ultraviolet (UV) cutoff scale of the underlying theory is identified with
the reduced Planck mass $m_P \simeq 2.4\times 10^{18}$ GeV.
As we will show, the potential for smooth hybrid inflation based on a
minimal K\"ahler potential does not realize inflation
with sub-Planckian values of the field. However, by employing a
non-minimal K\"ahler potential, one can realize inflation with
predictions that are consistent with the WMAP7 data.
We obtain in this case a scalar spectral index $n_s$
within the WMAP7 1-$\sigma$ bounds, and a large tensor to scalar ratio
(canonical measure of gravity waves) $r \lesssim 0.02$. The parameter region explored in this report
is expected to be tested soon by the Planck surveyor.

The simplified smooth inflation is defined by
the superpotential $W$,
\begin{equation}
W = S \left( \mu^2 -
\frac{(\Phi \overline{\Phi })^2}{m_P^2} \right)\,,
\label{superpot}
\end{equation}
where $S$ is a gauge singlet superfield, $\Phi$ and $\overline{\Phi }$
are a conjugate pair of superfields transforming as
nontrivial representations of some gauge group $G$, and $\mu$
is a superheavy mass. 
Note that in the expression for $W$ the UV cutoff $m_P$ 
(reduced Planck mass) has replaced the cutoff $M_*$ normally 
employed in smooth hybrid inflation models.
Both $W$ and $S$ carry the same $R$-charge,
while the combination $\Phi \overline{\Phi }$ is neutral under $U(1)_R$.
In addition, $W$ respects a $Z_2$ symmetry under which $S$ is even and
the combination $\Phi \overline{\Phi }$ is odd.
Thus, $W$ is the most general superpotential with leading order
non-renormalizable term which is consistent with the $R$, $Z_2$
and gauge symmetries.

\begin{figure}[th]
\centering \includegraphics[width=10cm]{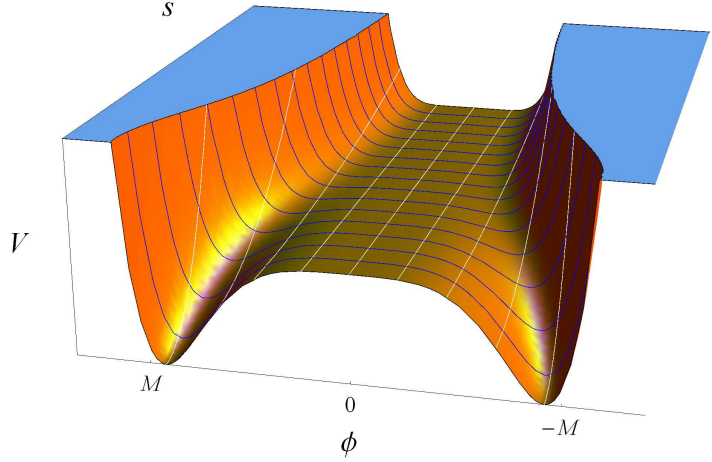}
\caption{The scalar potential $V$ of global susy smooth hybrid inflation
as a function of $\phi$ and $s$.}
\label{fig1}
\end{figure}

The SUGRA scalar potential is given by
\begin{equation}
V_{F} = \left. e^{K/m_{P}^{2}}\left(
K_{ij}^{-1}D_{Z_{i}}WD_{Z^{*}_j}W^{*}-3m_{P}^{-2}\left| W\right| ^{2}\right)
\right|_{Z_i = z_i},
\label{VF}
\end{equation}
with $z_{i} \in \{s,\,\phi,\,\overline{\phi },\cdots\}$ being the bosonic
components of the superfields
$Z_{i}\in \{S,\,\Phi,\,\overline{\Phi },\cdots\}$, and we have defined
\begin{eqnarray*}
D_{Z_{i}}W &\equiv &\frac{\partial W}{\partial Z_{i}}+m_{P}^{-2}\frac{%
\partial K}{\partial Z_{i}}W,
\,\,\,\,\,\, K_{ij} \equiv \frac{\partial ^{2}K}{\partial Z_{i}\partial Z_{j}^{*}},
\end{eqnarray*}
$D_{Z_{i}^{*}}W^{*}=\left( D_{Z_{i}}W\right) ^{*}$.
The minimal K\"ahler potential can be expanded as
\beq
K  =  |S|^{2}+ |\Phi|^{2} + |\overline{\Phi}|^{2}.  \label{kahler1}
\eeq
In the D-flat direction ($\overline{\phi}^{*} = \phi$),
and using Eqs. (\ref{superpot}, \ref{kahler1}) in Eq. (\ref{VF}), we obtain
\beq
V = \mu^4 \left[ \left(1 - \frac{|\phi|^4}{M^4}\right)^2
+ 8\,\frac{|s|^2\,|\phi|^6}{M^8}
+ \cdots \right],
\eeq
where $M = \sqrt{\mu\,m_P}$ is the vacuum expectation value (vev) of $\phi$
at the global SUSY minimum
($s = 0$, $\langle \phi \rangle = M $).
This potential is displayed in Fig.~\ref{fig1} which shows two valleys of minima $y_{\pm}(x)$
given in the large $x$ limit by
\beq
y_{\pm} \equiv \pm \sqrt{\sqrt{1+(3\,x^2)^2}-3\,x^2}
\approx \pm \frac{1}{\sqrt{6}\,x},
\eeq
where $y \equiv |\phi|/M$ and $x \equiv |s|/M$.

During inflation ($y = y_{+}$ and $x \gg 1$), and excluding SUGRA corrections,
the potential is given by,
\beq
V \simeq \mu^4 \left( 1 - \frac{1}{54\,x^4} \right).
\eeq
Using (leading order) slow-roll approximation, the scalar spectral
index $n_s$, the tensor to scalar ratio $r$, and the
running of the scalar spectral index $dn_s/d\ln k$ are given by
\bea
n_s &\simeq& 1+2\,\eta-6\,\epsilon
\simeq 1 - \frac{5}{3\,N_0}  \\
r &\simeq& 16\,\epsilon \simeq
\frac{8(2\pi\Delta_{\mathcal{R}})^{2/5}}{27N_0^2}  \\
\frac{d n_s}{d\ln k} &\simeq& 16\,\epsilon\,\eta
-24\,\epsilon^2 - 2\,\xi^2 \simeq - \frac{5}{3\,N_0^2}.
\eea
Here,
\beq
\epsilon = \frac{1}{4}\left( \frac{m_P}{M}\right)^2
\left( \frac{V'}{V}\right)^2, \,\,\,
\eta = \frac{1}{2}\left( \frac{m_P}{M}\right)^2
\left( \frac{V''}{V} \right),
\xi^2 = \frac{1}{4}\left( \frac{m_P}{M}\right)^4
\left( \frac{V' V'''}{V^2}\right), \,\,\,
\eeq
$N_0$ is the number of e-folds during inflation,
\beq
N_0 = 2\left( \frac{M}{m_P}\right) ^{2}\int_{x_e}^{x_{0}}\left( \frac{V}{%
V'}\right) dx,
\eeq
and the amplitude of the curvature perturbation is given by
\beq \label{perturb}
\Delta_{\mathcal{R}}^2 = \frac{1}{24\,\pi^2}
\left. \left( \frac{V/m_P^4}{\epsilon}\right)\right|_{x = x_0},
\eeq
where $\Delta_{\mathcal{R}}^2 = (2.43\pm 0.11)\times 10^{-9}$ is
the WMAP7 normalization at $k_0 = 0.002\, \rm{Mpc}^{-1}$ \cite{Komatsu:2010fb}.
The quantity $x_0$ denotes the field value at the pivot scale $k_0$, and
$x_e$ is the field value at the
end of inflation, defined by $|\eta(x_e)| = 1$. For $N_0 = 50$,
we obtain $n_s \simeq 0.968$, $r \simeq 3 \times 10^{-6} $
and $d n_s / d\ln k \simeq -7 \times 10^{-4}$, with
$x_0 \simeq 5$ and $M \simeq 6 \times 10^{16}$ GeV.

Including SUGRA corrections \cite{Linde:1997sj} with minimal (canonical) K\"ahler potential
modifies the above potential,
\beq
V_{\mathrm{SUGRA}} \simeq \mu^4 \left( 1 - \frac{1}{54\,x^4} +
\left( \frac{M}{m_P}\right)^2 \frac{8}{54\,x^2}
+  \left( \frac{M}{m_P}\right)^4 \frac{x^4}{2} \right).
\eeq
Taking values of $x_0$ and $M$ extracted from the global susy potential
one easily checks that the sugra corrections dominate the
global susy part. Thus,
sugra corrections can be expected to significantly alter the predictions of
$n_s$ and $r$ found earlier.
In  `simplified smooth hybrid inflation' ($M_* = m_P$)
with minimal (canonical) K\"ahler potential,
these sugra corrections require transplanckian field values
corresponding to $50$-$60$ e-folds of inflation. However, this requirement invalidates
the sugra expansion itself.
This is in contrast to `standard smooth hybrid inflation'
where the cutoff scale $M_*$ is allowed to vary below
$m_P$. Thus by suppressing sugra corrections somewhat
we can obtain values of $n_s $ just inside the WMAP7
2-$\sigma$ bounds, although with tiny values of $r$.

In order to obtain WMAP7 consistent red-tilted spectrum ($n_s \simeq 0.97$)
with observable values of $r$ in the simplified smooth hybrid inflation,
we consider, following Refs.~\cite{Shafi:2010jr,Civiletti:2011qg},
a non-minimal K\"ahler potential.
 [For `regular and standard smooth hybrid inflation' with non-minimal K\"ahler potential
see Refs.~\cite{BasteroGil:2006cm,urRehman:2006hu}]. The K\"ahler potential with non-minimal terms is given by,
\bea    \label{kahler2}
K  &=&  |S|^{2}+ |\Phi|^{2} + |\overline{\Phi}|^{2} \nonumber \\
&&  + \kappa_S \frac{|S|^4}{4\,m_P^2}
+ \kappa_{\Phi} \frac{|\Phi|^4}{4\,m_P^2}
+ \kappa_{\overline{\Phi}} \frac{|\overline{\Phi}|^4}{4\,m_P^2}
+ \kappa_{S\Phi} \frac{|S|^2|\Phi|^2}{m_P^2}
+ \kappa_{S\overline{\Phi}} \frac{|S|^2|\overline{\Phi}|^2}{m_P^2}
+ \kappa_{\Phi\overline{\Phi}} \frac{|\Phi|^2|\overline{\Phi}|^2}{m_P^2}
\nonumber  \\
&& + \kappa_{SS} \frac{|S|^6}{6\,m_P^4} + \cdots \,.
\eea
The corresponding scalar potential takes the following form,
\beq
V \simeq \mu^4 \left( 1 - \frac{1}{54\,x^4} +
\left(-\kappa_S\,x^2 + \frac{8+3\,\kappa_S}{54\,x^2} \right)
\left( \frac{M}{m_P}\right)^2
+ \gamma_S\left( \frac{M}{m_P}\right)^4 \frac{x^4}{2}  \right),
\eeq
where $\gamma _{S}=1-\frac{7\kappa _{S}}{2}+2\kappa _{S}^{2}-3\kappa _{SS}$.
We have suppressed radiative corrections in the above potential since
there is no direct renormalizable coupling of the inflaton with the other fields.
We have also ignored the soft susy breaking terms as their contribution will be
negligible in the parameter range consistent with the WMAP7 (2$\sigma$) bounds 
\cite{Senoguz:2004vu,Rehman:2009nq}.

\begin{figure}[t]
\centering \includegraphics[width=7.5cm]{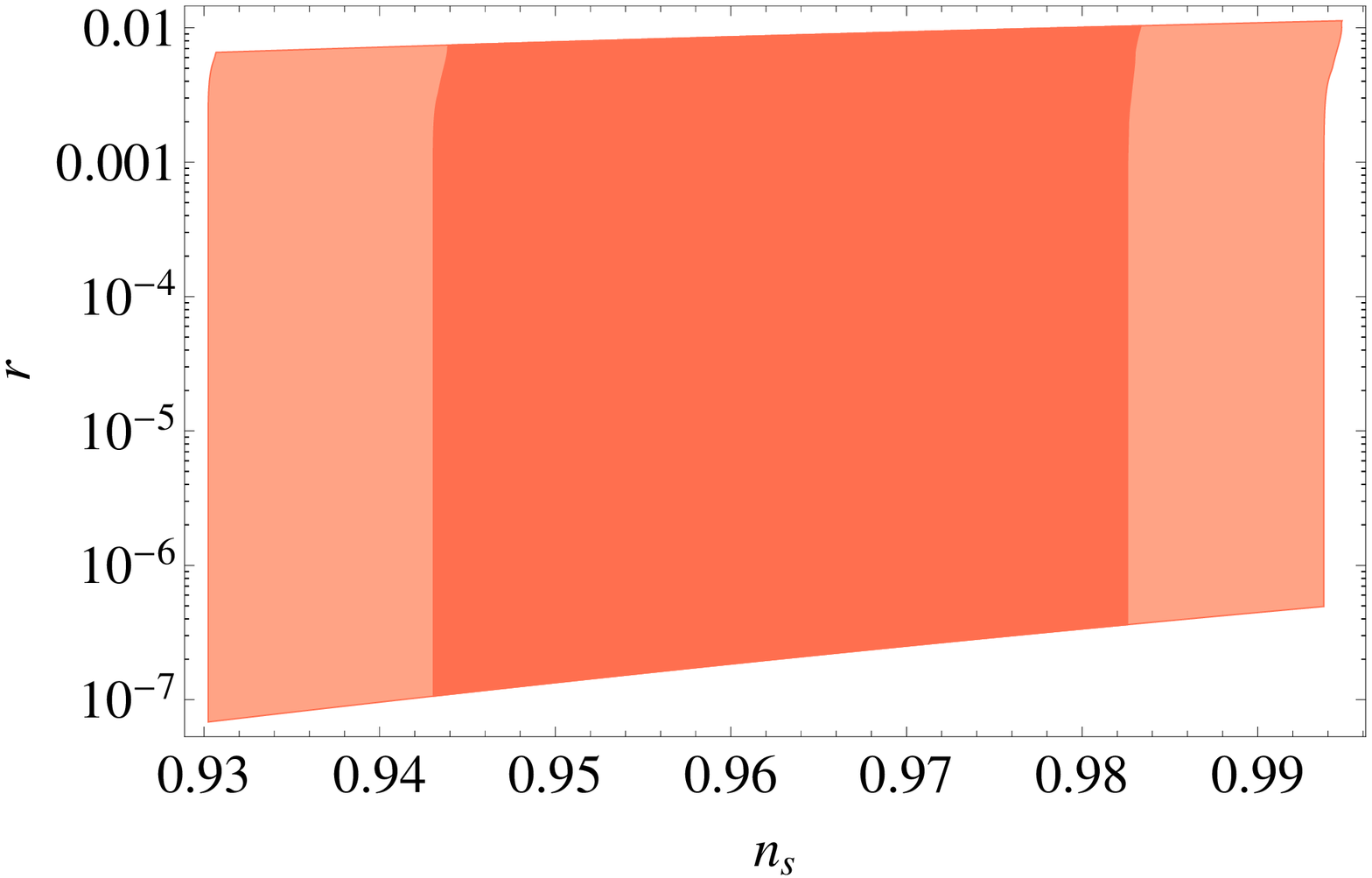}
\centering \includegraphics[width=7.5cm]{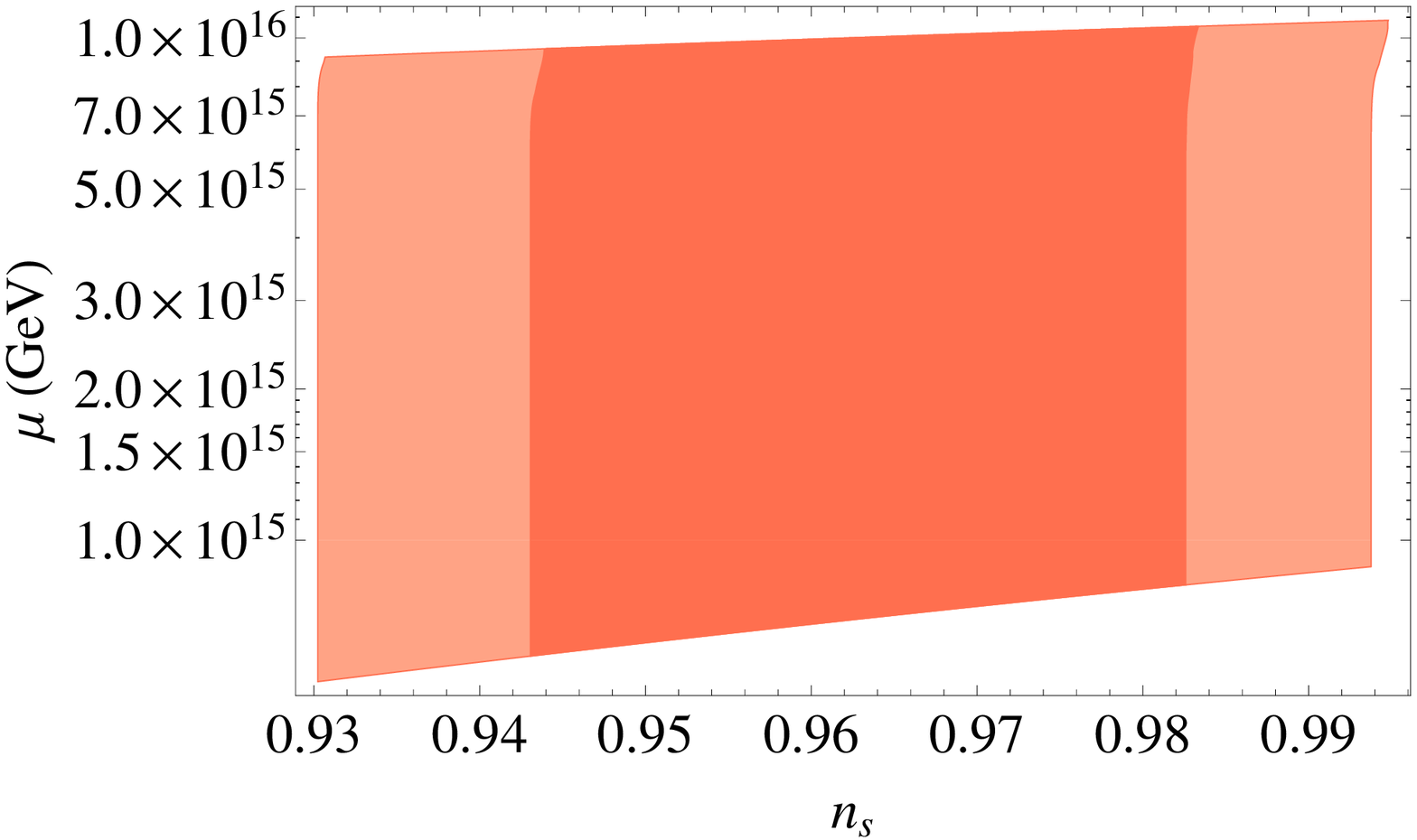}
\caption{$r$ (left panel) and $\mu$ (right panel)
vs $n_s$ with $N_0 = 50$.
The WMAP7 1-$\sigma$ and 2-$\sigma$ bounds are shown in the
dark and light red regions respectively.
The upper and lower boundary curves for $r$ and $\mu$ represent
the $|s_0| = m_P$ and $\kappa_{SS}=-1$ constraints, respectively.}
\label{fig2}
\end{figure}

\begin{figure}[t]
\centering \includegraphics[width=7.5cm]{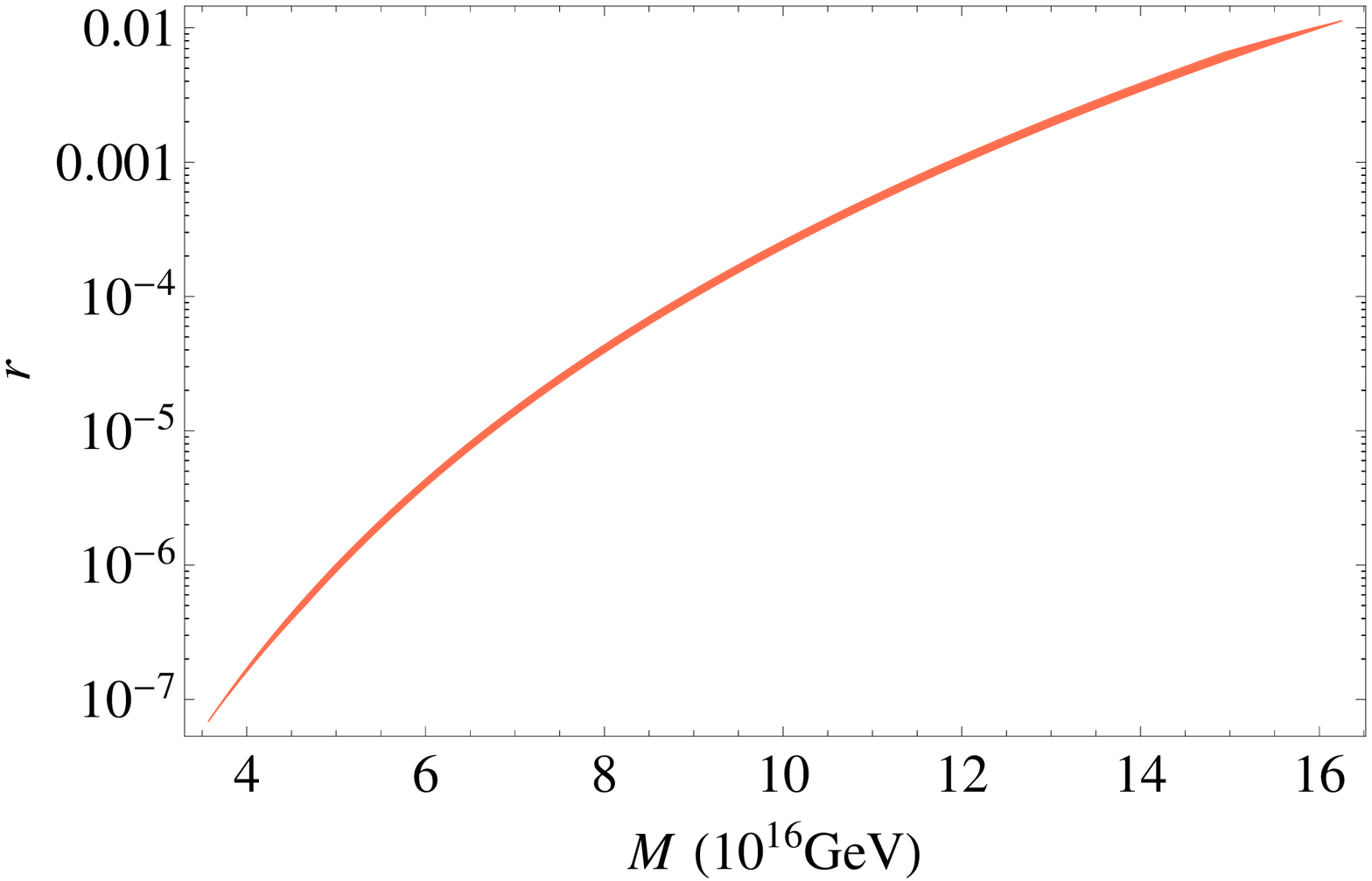}
\centering \includegraphics[width=7.5cm]{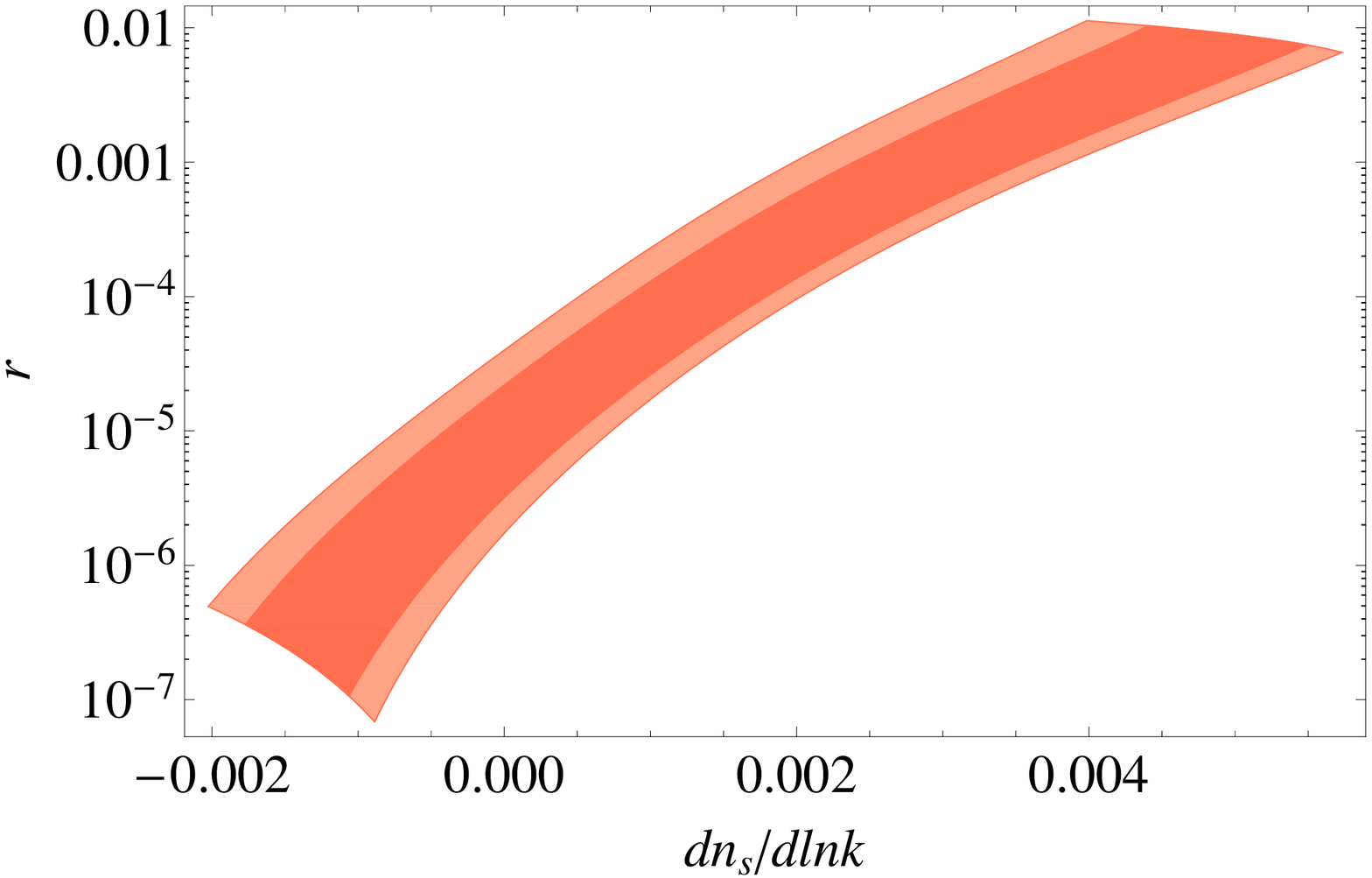}
\caption{Behavior of $r$ with respect to $M$ (left panel) and
$d\ln n_s / dk$ (right panel) for $N_0 = 50$.
The WMAP7 1-$\sigma$ and 2-$\sigma$ bounds are shown in the
dark and light red regions respectively.}
\label{fig3}
\end{figure}

\begin{figure}[t]
\centering \includegraphics[width=7.5cm]{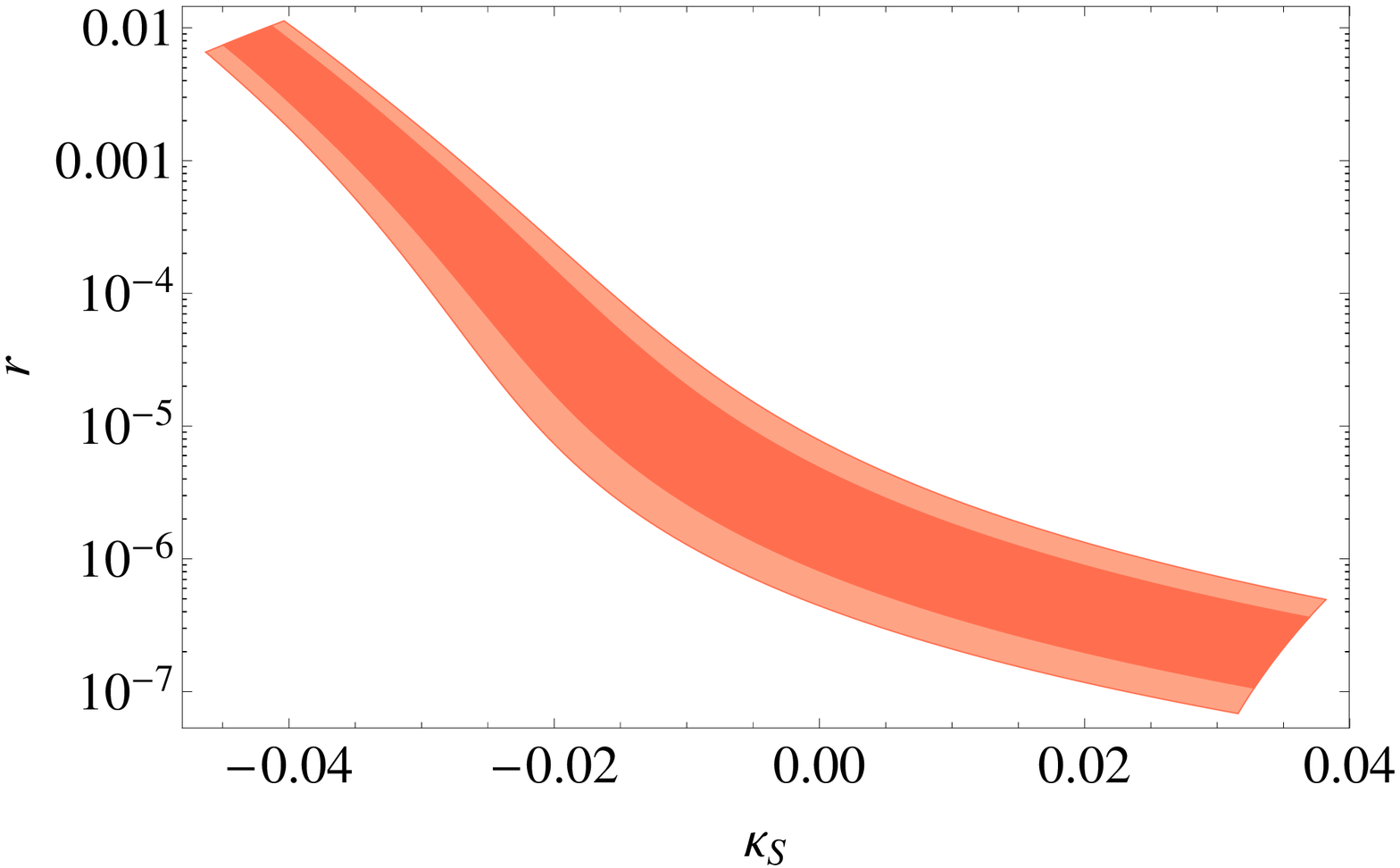}
\centering \includegraphics[width=7.5cm]{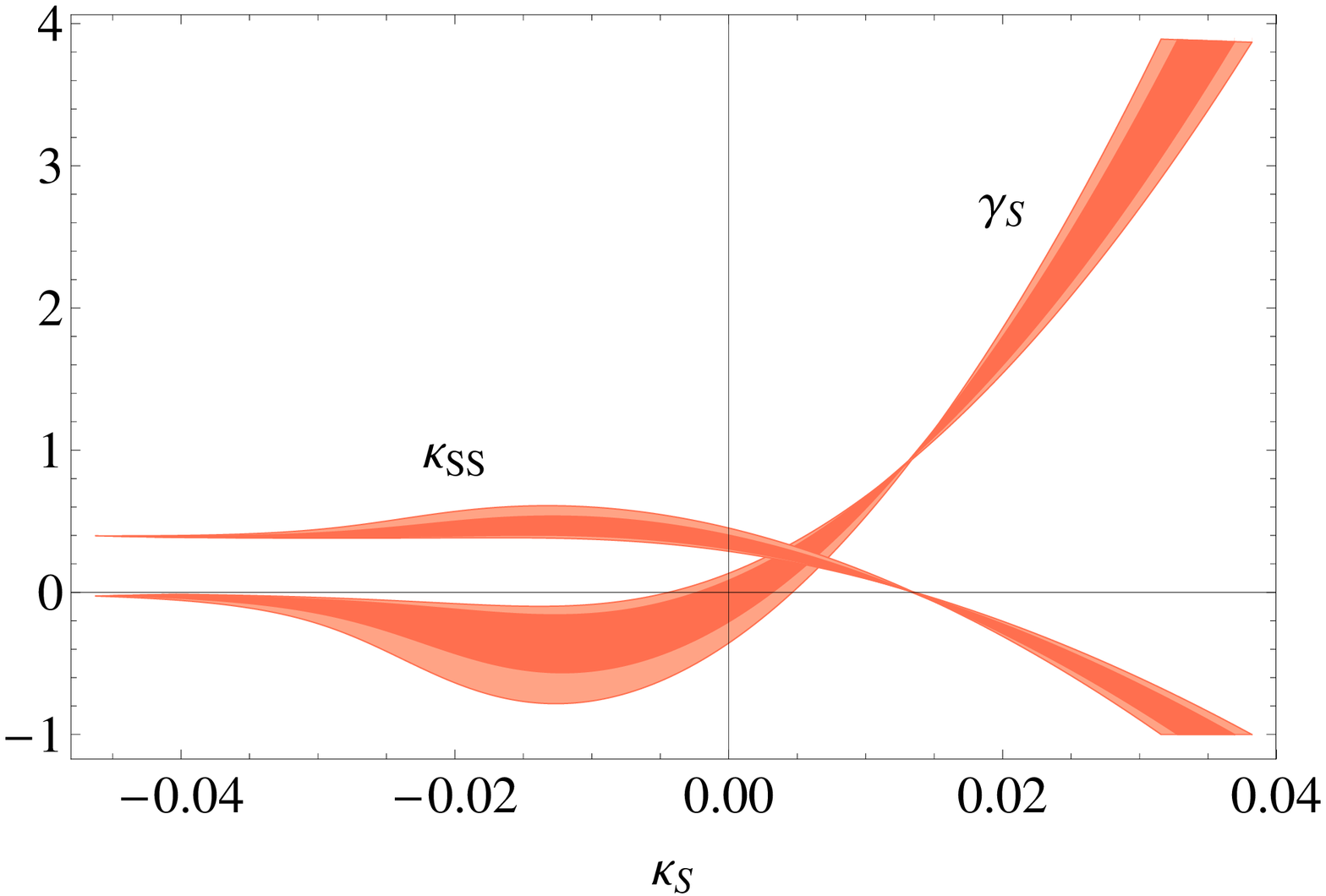}
\caption{$r$, $\gamma_S$ and $\kappa_{SS}$ as a function of $\kappa_S$ with $N_0 = 50$.
The WMAP7 1-$\sigma$ and 2-$\sigma$ bounds are shown in the
dark and light red regions respectively.}
\label{fig4}
\end{figure}

The predictions for the various inflationary parameters are
obtained by employing the slow-roll approximation
and are displayed in Figs.~(\ref{fig2}-\ref{fig4}).
To achieve better precision in the numerical results,
we have also included the next-to-leading order corrections
\cite{Stewart:1993bc,NeferSenoguz:2008nn} in the slow roll expansion for the
quantities $n_s$, $r$, $d n_s /d\ln k$, and $\Delta_{\mathcal{R}}$.
We require $(|\kappa_S|,\,|\kappa_{SS}|)\leq1$ and $|s_0| \leq m_P$.
In Fig.~(\ref{fig2}) we have presented the behavior of $r$ and $\mu$
with respect to $n_s$ along with the WMAP7 1-$\sigma$
and 2-$\sigma$ bounds.
The upper bound on $r$ comes from the constraint $|s_0| \leq m_P$,
 whereas the lower boundary curve represents
the $\kappa_{SS}=-1$ constraint.
These plots show that with the help of non-minimal K\"ahler potential,
$r$ can be increased by up to four orders of magnitudes as compared to its value
from the global susy potential.
The large $r$ solutions, however, require values of $M$ larger than
the grand unified theory (GUT) scale $\sim 2 \times 10^{16}$
(see Fig.~(\ref{fig3})). From Eq.~(\ref{perturb}) and the definition
$M \equiv \sqrt{\mu \, m_P}$, one finds the following
approximate relation,
\beq
r \simeq  \left(\frac{2}{3\, \pi^2 \, \Delta_{\mathcal{R}}^2} \right)
\left( \frac{M}{m_P} \right)^8
= \left( \frac{M}{3.35\times 10^{16}\,GeV}\right)^4
\left( \frac{M}{m_P}\right)^4.
\eeq
This relation provides a reasonable estimate of the otherwise more accurately
calculated numerical result displayed in Fig.~(\ref{fig3}). However,
this relation alone is insufficient to explain the upper bound on the values of $r$
which is discussed below in some detail. Furthermore, the values of $dn_s/d\ln k$
shown in the right panel of Fig.~(\ref{fig3}) are reasonably small and
in accord with the WMAP7 data assumptions.

In Fig.~(\ref{fig4}), left panel, we show how $r$ varies with respect to $\kappa_S$,
while the right panel shows the relationship between $\kappa_{SS}$ and $\gamma_S$ to $\kappa_S$.
As anticipated, the minimal case with $(\kappa_S,\,\kappa_{SS})=(0,0)$
is not consistent with the WMAP7 1-$\sigma$ and 2-$\sigma$ bounds.
Moreover, in agreement with earlier observations
(Refs.~\cite{Shafi:2010jr,Civiletti:2011qg,Rehman:2009wv}),
large (observable) $r$ solutions are obtained with the potential form,
$V/\mu^4 \simeq 1+$ quadratic - quartic, with
$(\kappa_S,\,\gamma_S)<(0,0)$, as shown in
the right panel of Fig.~(\ref{fig4}).
This behavior can be explained with the requirements $V'(x_0)>0$ and
$V''(x_0)<0$ (or $n_s < 1$), with $\epsilon(x_0) \ll \eta(x_0)$ at the pivot scale.
Consider the following approximate form of $\epsilon$ and $\eta$,
\beq
\epsilon \simeq f^2\left( \frac{(M/m_P)^4}{27\,f^6}
-\kappa_S+\gamma_S\,f^2\right)^2, \,\,\,
\eta \simeq \frac{5\,(M/m_P)^4}{27\,f^6} -\kappa_S + 3\,\gamma_S\,f^2,
\eeq
where, $f \equiv |s|/m_P$. In the large $r$ limit,
the contribution from the global susy part of the potential
is negligible at the pivot scale, and it becomes important only near
the end of inflation. After neglecting this contribution,
one can check that the choice $(\kappa_S,\,\gamma_S)<(0,0)$
is the only possibility which is consistent with large
values of $r$ and a red-tilted spectrum $n_s < 1$.

\begin{figure}[t]
\centering \includegraphics[width=7.5cm]{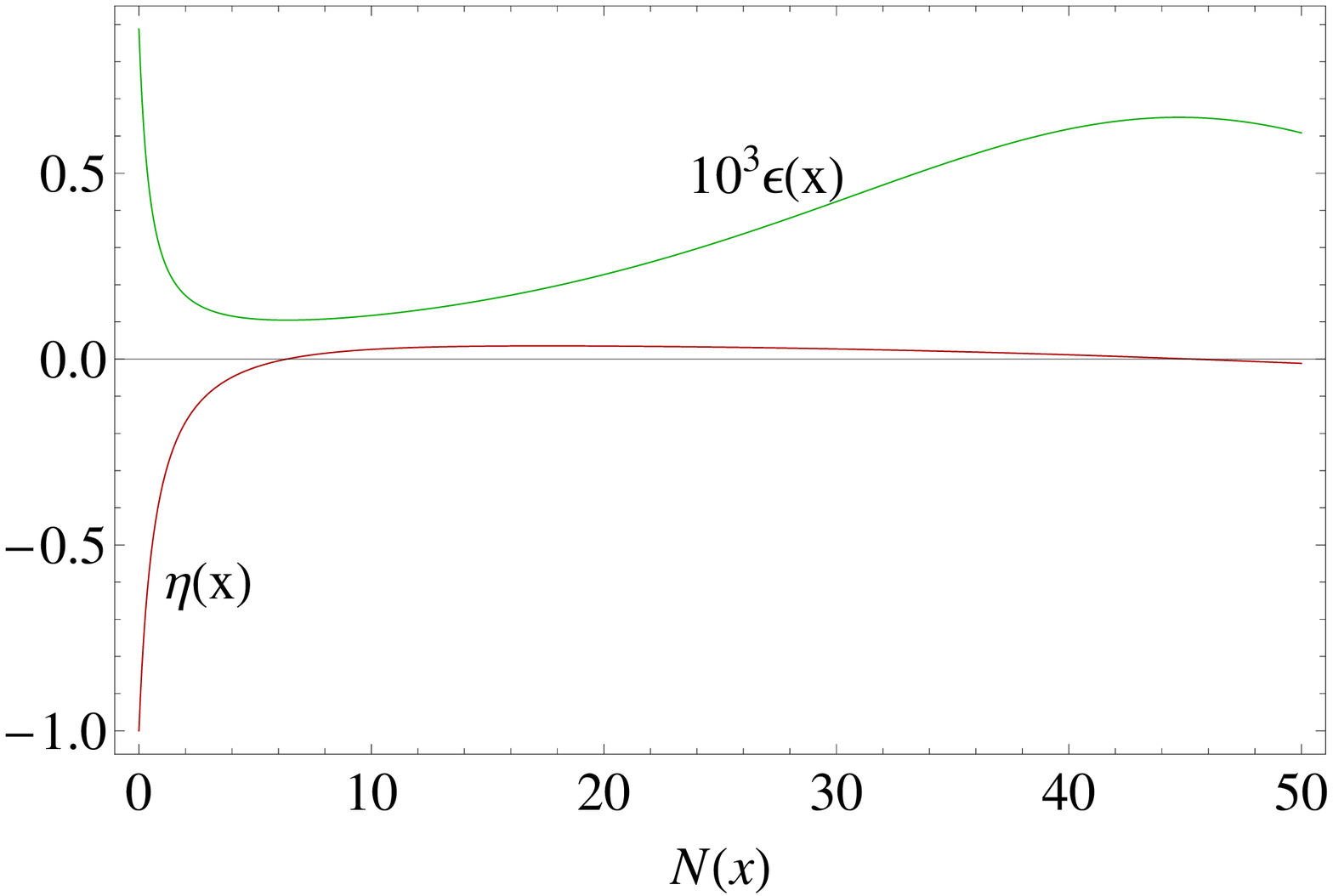}
\centering \includegraphics[width=7.5cm]{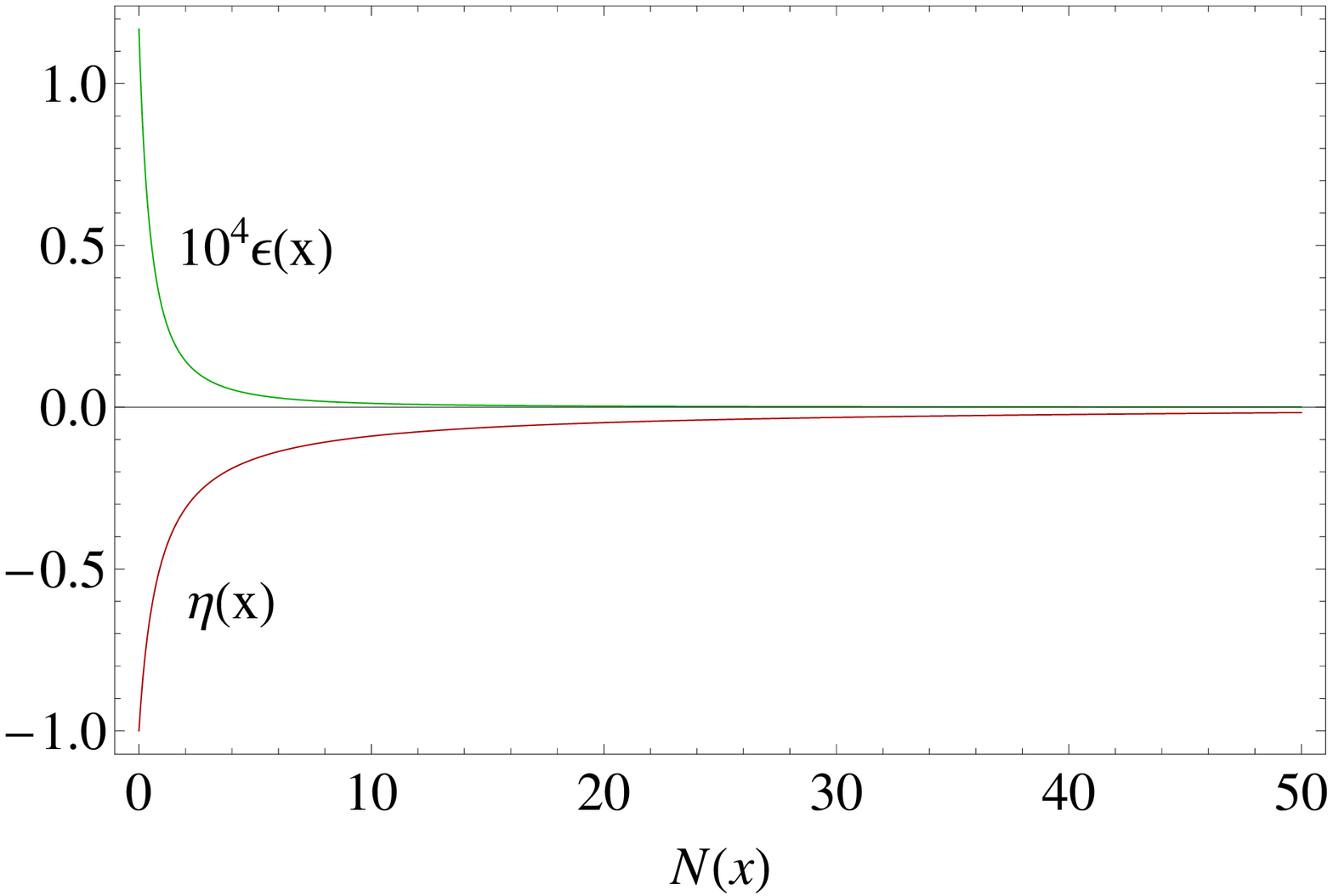}
\caption{$\epsilon(x)$ and $\eta(x)$ as function of $N(x)$ for
$r = 10^{-2}$ (left panel) and $r = 10^{-6}$ (right panel), with
$n_s = 0.968$ and $N_0 = 50$.}
\label{fig5}
\end{figure}

Next we turn our attention to the explanation of the upper bound on $r$.
It might be tempting to justify the observed upper limit on $r$
through the well known bound \cite{Lyth:1996im},
\beq \label{lyth}
r \lesssim 0.006 \left( \frac{50}{N_0}\right)^2
\left( \frac{\Delta s}{m_P}\right)^2,
\eeq
which is derived with the assumption of a monotonically increasing
$\epsilon$ during inflation. With $N_0 = 50$ and
$s_0 = m_P$, the bound in eq. (\ref{lyth}) predicts $r \lesssim 0.006$, 
which is in apparent contradiction
with our result $r \lesssim 0.02$. Actually, the assumption
of monotonically increasing $\epsilon$ breaks down for the large $r$ solutions,
as shown explicitly in Fig.~(\ref{fig5}) for the central value of
the scalar spectral index $n_s = 0.968$. (Also, see
Refs.~\cite{BenDayan:2009kv,Hotchkiss:2011gz}
for large $r$ solutions with small field excursions
in the context of this bound.).

Let us consider the following relation for the
variation of $\epsilon$,
\beq
\epsilon'(x) = 2\left( \frac{M}{m_P} \right)
\sqrt{\epsilon} \, (\eta - 2\,\epsilon).
\eeq
During inflation $\eta$ remains dominant over $\epsilon$
and controls the evolution of $\epsilon$. For large $r$ solutions
inflation starts and ends with $\eta < 0$ (recall that $\eta(x_0)<0$
is required for red-tilted spectrum and $\eta(x_e)=-1$), while passing through
the inflection point $\eta = 0$. The change in the sign of $\eta$
actually introduces the non-monotonic behavior of $\epsilon$
as shown in Fig.~(\ref{fig5}). Therefore, Eq.~(\ref{lyth})
underestimates the upper bound on $r$ in this case.

It is interesting to note that the small $r$ solutions
do exhibit a monotonic behavior of $\epsilon$
(see the right panel of Fig.~(\ref{fig5})).
This comes from  the fact that the
small $r$ solutions favor $(\kappa_S,\,\gamma_S)>(0,0)$, since
large values of the field generate a blue-tilted spectrum $n_s>1$
caused by positive values of $\eta(x_0)>\epsilon(x_0)> 0$.
Therefore, during inflation the quartic term remains
sub-dominant and this makes $\eta$ negative and $\epsilon$
monotonically increasing.

In order to provide a semi-analytical estimate for the upper bound
on $r$, we employ Eqs.~(16-17) with $f_0 = 1$ to approximate
the number of e-folds,
\beq
N_0 \simeq \frac{1}{-6\,\kappa_S} \ln \left[\frac{27}{20}
\left( \frac{m_P}{3.35 \times 10^{16}\, GeV} \right)^2
\frac{-\,\kappa_S^3}{\left(-\kappa_S + \gamma_S \right)^4} \right],
\eeq
with,
\beq
\gamma_S \simeq - \left( \frac{1 - n_s -4\,\kappa_S }{6} \right),
\,\,\, r \simeq 16\,(-\kappa_S + \gamma_S)^2.
\eeq
Now, for a given value of $n_s$ and $N_0$, one can calculate the values of
$\kappa_S$, $\gamma_S$ and $r$. For $n_s = 0.968$ we obtain $\kappa_S \simeq -0.047$,
$\gamma_S \simeq -0.026$ and $r \simeq 0.011$,
which is in good agreement with our numerical results.

 To summarize, we have considered a simplified version of smooth hybrid
inflation with a single UV cutoff $m_P$. With minimal K\"ahler potential,
the presence of sugra corrections invalidates the otherwise successful inflation
obtained with the global susy part of the potential. However, with a non-minimal
extension of the K\"ahler potential we obtain a red-tilted spectrum
consistent with the WMAP7 data, and we also achieve up to four orders of
magnitude increase in the value of $r (\sim 0.02)$, compared to the global susy result
$r \sim 10^{-6}$. These large $r$ solutions can be expected to be observed by
the PLANCK satellite.

\section*{Acknowledgments}
This work is supported in part by the DOE under grant \# DE-FG02-91ER40626
(M.R. and Q.S.), by the Bartol Research Institute (M.R. and Q.S.)
and by the Department of Physics, University of Basel.

\end{document}